\documentclass[%
 reprint,
 superscriptaddress,
 amsmath,amssymb,
 aps,
 pra,
]{revtex4-1}

\usepackage{graphicx}
\usepackage{dcolumn}
\usepackage{bm}
\usepackage{braket}
\usepackage{hyperref}
\usepackage{tikz}
\definecolor{myblue}{RGB}{0,102,204}
\definecolor{myorange}{RGB}{255,128,0}

\begin{document}

\preprint{APS/123-QED}

\title{Time-Reversed Task Events and Thermodynamic Bounds for Stochastic Thermal Machines}

\author{Kaiyuan Cao}
 \email{kycao@yzu.edu.cn}
\affiliation{College of Physics Science and Technology, Yangzhou University, Yangzhou 225009, People's Republic of China}

\author{Bei Xiao}
\affiliation{College of Physics Science and Technology, Yangzhou University, Yangzhou 225009, People's Republic of China}

\author{Jian Wang}
 \email{phcwj@hotmail.com}
\affiliation{College of Physics Science and Technology, Yangzhou University, Yangzhou 225009, People's Republic of China}

\date{\today}

\begin{abstract}
    We develop a task-based framework that relates finite-time performance to thermodynamic irreversibility in stochastic heat engines and refrigerators. For tasks defined by reaching an accumulated-work threshold within a prescribed time, we construct the reverse event through exact trajectory time reversal. The binary relative entropy between forward and reverse task probabilities provides a lower bound on mean entropy production. Combined with the first law and entropy balance, this bound constrains irreversible work loss in engines and excess input work in refrigerators. We illustrate the framework with a three-level stochastic system and test the bounds against its full dynamics. These results show how success-or-failure statistics alone can reveal irreversibility and constrain the energetic costs of finite-time thermal tasks. We also outline an experimental route to testing the framework using a monitored three-level system coupled to engineered thermal reservoirs and a work channel.
\end{abstract}

\maketitle


{\it Introduction.---}
Heat engines and refrigerators mediate energy conversion between work and heat reservoirs in opposite directions. Their reversible performance in bounded by the Carnot efficiency and the Carnot coefficient of performance, respectively, whereas finite-time operation generally entails dissipation. Understanding the resulting performance loss connects classical finite-time thermodynamics with the stochastic dynamics of microscopic thermal mechanics \cite{Benenti2017pr, Seifert2012rpp, Ciliberto2017prx}. This connection has motivated extensive research on efficiency at maximum power \cite{Schmiedl2008epl, Esposito2010prl}, universal trade-offs between power and efficiency \cite{Shiraishi2016prl}, and the joint constraints on power, efficiency, and output fluctuations \cite{Pietzonka2018prl}. Geometric approaches further relate finite-time dissipation to the paths followed by thermodynamic control parameters, providing bounds and optimization principles for microscopic heat engines \cite{Brandner2020prl, Frim2022prl}. Together, these advances establish entropy production as a central quantity linking microscopic irreversibility to the macroscopic costs of producing work or maintaining refrigeration.

For a heat engine, this cost appears as lost work relative to the Carnot benchmark; for a refrigerator, it appears as additional input work beyond the reversible requirement for extracting a prescribed amount of heat from the cold reservoir. Although these costs are governed by total entropy production, its direct evaluation can require information about heat exchanges, transition rates, or complete trajectories that is unavailable experimentally. The relation between dissipation and statistical distinguishability of forward and time-reversed dynamics provides a route to inference from incomplete observations \cite{Kawai2007prl, Gomez2008pre, Roldan2010prl}. Complementary approaches use current fluctuations and thermodynamic uncertainty relations to constrain dissipation \cite{Barato2015prl, Gingrich2016prl}, while thermodynamic inference more broadly seeks to extract such constraints from accessible dynamical data \cite{Seifert2019arcmp}. Recent advances have extended this program to partially accessible networks and transition-based waiting-time distributions \cite{van2022prx}, time-resolved trajectory snippets \cite{van2023prl}, general observed statistics \cite{Nitzan2023prr}, and fluctuating entropy production at a coarse-grained level \cite{Degunther2024prr}. Related studies address periodically driven systems \cite{Maier2024jsp}, continuous dynamics \cite{Meyberg2024jpa}, and the effects of finite measurement resolution and finite statistics \cite{Fritz2025pre}. These developments motivate a particularly restricted observation scheme: retaining only whether a prescribed operational task has been completed within a given observation time.

A natural instance is stochastic work extraction, where success is defined by an accumulated work current reaching a prescribed threshold before a fixed deadline. Such tasks connect thermal machine performance to first-passage theory, extreme-value statistics, and stopping-time thermodynamics \cite{Gingrich2017prl, Neri2017prx, Neri2019jsm, Neri2020prl}. Previous studies have established thermodynamic constrains on current first-passage times \cite{Gingrich2017prl}, derived universal properties of entropy-production extreme \cite{Neri2017prx}, formulated fluctuation relations at stopping times \cite{Neri2019jsm, Neri2020prl}, and used first-passage observables to estimate entropy production \cite{Neri2022jpa, Manzano2022pre}. Despite these advances, it remains unclear how irreversibility constrains the probability of meeting a prescribed performance target within a finite time for engines and refrigerators operating under stochastic conditions. 

Addressing this question requires relating task success to dissipation while accounting for fluctuations in the trajectories. We consider a setting in which each run is classified only as a success or failure: this provides a minimal performance measure, but discards the details trajectory and, for a threshold task, the actual first-passage time. Moreover, the reverse probability cannot in general be defined by imposing the same threshold condition on the reversed accumulated current. It must correspond to the time-reversed image of full forward task event, which, for a running-threshold event, generally involves a terminal drawdown of the reversed current.

Specifically, let $q_{F}(t)$ denote the probability of completing the forward task by time $t$, and $q_{B}(t)$ denote the probability of its exact time-reversed image under the reverse dynamics. Mapping each trajectory to a binary outcome cannot increase the relative entropy between the forward and time-reversed path measures. Under the standard identification of this path-space relative entropy with the mean total entropy production, the data-processing inequality gives
\begin{equation}
    \langle \Sigma_{t} \rangle \geq k_{B} D_{\rm bin}\!\left(q_F(t)\middle\|q_B(t)\right).
\end{equation}
The distinguishability of the forward and reverse task outcomes therefore provides a lower bound on the mean dissipation that can be inferred from the success probabilities alone, without reconstructing the full stochastic trajectories or calculating the entropy production along each trajectory.

Combing this information bound with the first law and the entropy balance translates task statistics into constraints on thermal machine performance. For a heat engines operating between reservoirs at temperatures $T_{H}>T_{C}$, the bound takes the form
\begin{equation}
    \langle W_{\mathrm{out}}\rangle \leq \eta_{C}\langle Q_{H}\rangle - k_{B}T_{C}D_{\mathrm{bin}}\big(q_{F}(t)\Vert q_{B}(t)\big).
\end{equation}
Here, $\eta_{C}=1-\frac{T_{C}}{T_{H}}$ is the Carnot efficiency, $\langle Q_{H}\rangle$ is the mean heat absorbed from the hot reservoir, and $\langle W_{\mathrm{out}}\rangle$ is the mean work output. The left-hand side measures the work lost relative to reversible conversion of the same mean heat input. Forward and reverse task outcomes that are statistically distinguishable therefore impose a positive minimum work loss.

For a refrigerator, the same bounds the work input required to extract a mean heat $\langle Q_{C}\rangle$ from a cold reservoir:
\begin{equation}
    \langle W_{\mathrm{in}}\rangle \geq \frac{\langle Q_{C}\rangle}{\mathrm{COP}_{C}} + k_{B}T_{H}D_{\mathrm{bin}}\big(q_{F}(t)\Vert q_{B}(t)\big).
\end{equation}
Here, $\mathrm{COP}_{C}=\frac{T_{C}}{T_{H}-T_{C}}$ is the Carnot coefficient of performance. The first term is the reversible work requirement for the specified mean cooling load, while the second gives a lower bound on the excess work inferred from task statistics. The engine and refrigerator bounds therefore have complementary interpretations: task distinguishability sets a minimum loss of work output for an engine and a minimum additional work input for a refrigerator. The respective conversion factors are $k_{B}T_{C}$ and $k_{B}T_{H}$; the full bounds also depend on the task probabilities and mean heat exchanges, which may vary with both reservoir temperatures.

Together, these results provide a framework for connecting the probability of completing a finite-time task to the energetic cost of operating a thermal machine. We illustrate the framework using a three-state stochastic model with work recorded along individual trajectories. By calculating the forward and reverse task probabilities, we examine how the observation time and task threshold affect the inferred dissipation and assess the tightness of the performance bounds against the full stochastic dynamics. This analysis establishes how much of the irreversible work loss or excess work input can be resolved when each trajectory is reduced to a single success-or-failure outcome.

\begin{figure}
    \centering
    \includegraphics[width=1\linewidth]{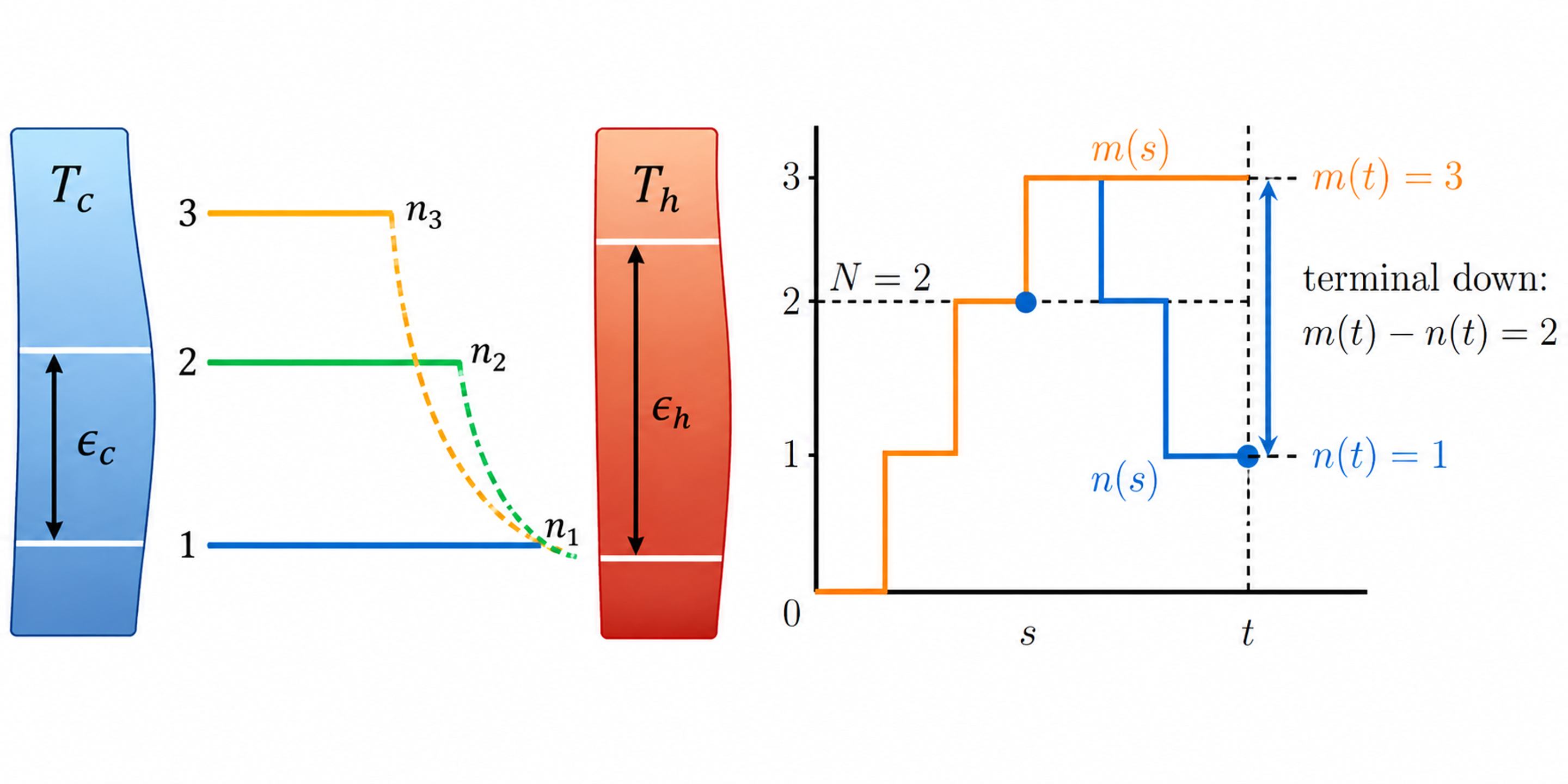}
    \caption{The left panel: Schematic of a three-level thermal machine coupled to hot and cold reservoirs. The right panel: Forward and strictly time-reversed task events for the same representative work-count trajectory. The forward count $n(s)$ and its running maximum $m(s)=\max_{0\leq u\leq s}n(u)$. The forward task is completed when $m(t)\geq N$. The exact reversed event of the forward task is the terminal drawdown event $\max_{0\leq r\leq t}\bar n(r)\geq N$, equivalently $m(t)-n(t)\geq N$. }
    \label{fig: three.level.machine}
\end{figure}

{\it Three-level thermal machine.---}
To illustrate the task-based bounds, we consider a three-state Markov jump model inspired by the three-level maser, a minimal paradigm for studying the thermodynamic behavior of thermal machines \cite{Scovil1959prl, Kosloff2014arpc, Correa2014sr, Uzdin2015prx}. The system has energies $E_{1}<E_{2}<E_{3}$ and is coupled to hot and cold reservoirs at temperatures $T_{H}>T_{C}$. The hot and cold reservoirs induce transitions $1 \leftrightarrow 3$ and $1 \leftrightarrow 2$, respectively, while an ideal work channel connects $2 \leftrightarrow 3$ (see the left panel in Fig.~\ref{fig: three.level.machine}). The corresponding energy quanta are $\epsilon_{H} = E_{3}-E_{1}$, $\epsilon_{C} = E_{2}-E_{1}$, and $W^{*} = \epsilon_{H}-\epsilon_{C}$.

The thermal rates obey local detailed balance \cite{Seifert2012rpp},
\begin{equation}\label{eq: ldb}
    \frac{k^{(\alpha)}_{i\to j}}{k^{(\alpha)}_{j\to i}} = \exp[-\beta_\alpha(E_j-E_i)], \quad \beta_\alpha=(k_BT_\alpha)^{-1}
\end{equation}
with $\alpha\in\{H, C\}$. We model the ideal work channel by symmetric transition rates, $k^{(W)}_{2\to3}=k^{(W)}_{3\to2}$. For symmetric work-channel rates, the machine operates as a heat
engine when $\beta_{C}\epsilon_{C}>\beta_{H}\epsilon_{H}$ and as a refrigerator when $\beta_{C}\epsilon_{C}<\beta_{H}\epsilon_{H}$ \cite{Scovil1959prl, Geva1992jcp}. In stochastic thermodynamics, the entropy transferred to the environment by a jump is given by the logarithm of the ratio of its forward and reverse rates \cite{Maes2021spln}. Thus, work-channel jumps exchange energy but contribute no entropy flow to the environment. The populations obey the Markov master equation $\dot{\boldsymbol{p}} = \mathsf{L}\boldsymbol{p}$ \cite{Esposito2010pre} with the generator $\mathsf{L}$. We initialize the system in its stationary distribution $\boldsymbol{p}^{\mathrm{ss}}$, which satisfies $\mathsf{L}\boldsymbol{p}^{\mathrm{ss}} = 0$.

We take heat to be positive when it enters the system and work to be positive when it is extracted, so that $\Delta E=Q_{H}+Q_{C}-W_{\mathrm{out}}$ along each trajectory.

{\it Forward and reverse task events.---}
For a trajectory $\omega_{t}$ observed during a fixed interval $[0, t]$, let $n(s)$ denotes the integrated work count, with opposite increments assigned to the reverse transitions. The forward task is the completion of $N > 0$ net work quanta at any time before $t$:
\begin{equation}\label{eq: forward.task}
    A_{N}(t) = \left\{ \max_{0\leq s\leq t}n(s)\geq N \right\}, \quad q_{F}(t) = P_{F}[A_{N}(t)].
\end{equation}
This is a fixed-time task event: once the threshold has been reached, the task remains completed even if the count subsequently decreases.

The strict reverse task is defined as the time-reversed image of the complete forward event,
\begin{equation}\label{eq: reverse.task.definition}
    \Theta A_{N}(t) = \{ \Theta \omega_{t}: \omega_{t}\in A_{N}(t) \}, \quad q_{B}(t) = P_{B}[\Theta A_{N}(t)].
\end{equation}
For the reversed trajectory, define
\begin{equation}
    \bar{n} = n(t-r)-n(t), \quad 0\leq r\leq t.
\end{equation}
The reverse event can therefore be written as
\begin{equation}\label{eq: reverse.task}
    \Theta A_{N}(t) = \left\{ \max_{0\leq r\leq t}\bar{n}(r)\geq N \right\}.
\end{equation}
Thus, the strict reverse task is a terminal drawdown of at least $N$ from the running maximum (see the right panel in Fig.~\ref{fig: three.level.machine}). It should not be identified with the fixed-boundary reverse first-passage event $\{\min_{0\leq r\leq t}\bar{n}(r)\leq -N\}$ \cite{Neri2017prx, Falasco2020prl, Pal2021prr}, which corresponds to a different operational task.

{\it Binary coarse-graining and entropy bound.---}
We now coarse-grain each complete trajectory to a binary task variable \cite{Cover2005, Marin2008pre, van2014ieee, Harunari2022prx}. For forward and backward paths, respectively, define \cite{Donati2017jcp, Guyader2020jsp}
\begin{equation}
    Y_{F}(\omega_{t}) = \mathbf{1}_{A_{N}(t)}(\omega_{t}), \quad Y_{B}(\omega_{t}) = \mathbf{1}_{\Theta A_{N}(t)}(\omega_{t}).
\end{equation}
Consequently, this deterministic coarse-graining induces the binary distributions
\begin{equation}
    P_{F}(Y_{F}=1) = q_{F}(t), \quad P_{B}(Y_{B}=1) = q_{B}(t).
\end{equation}
The data-processing inequality gives \cite{csiszar1967, Marin2008pre}
\begin{equation}\label{eq: task.DPI}
    D(P_{F}\Vert  P_B) \geq D_{\mathrm{bin}}\big(q_{F}(t)\Vert q_{B}(t)\big),
\end{equation}
where 
\begin{equation}
    D_{\mathrm{bin}}(p\Vert q) = p\ln{\frac{p}{q}} + (1-p)\ln{\frac{1-p}{1-q}}.
\end{equation}
For a trajectory $\omega_{t}$, the total entropy production is 
\begin{equation}
    \Sigma_{t}[\omega_{t}] = \Delta s_{\mathrm{sys}}[\omega_{t}] + \Delta s_{\mathrm{med}}[\omega_{t}],
\end{equation}
where $\Delta s_{\mathrm{sys}}$ denotes the stochastic system entropy \cite{Seifert2005prl, Seifert2012rpp, van2015pa}, and $\Delta s_{\mathrm{med}}$ is the medium contribution determined by the heat exchanges with reservoirs. Local detailed balance implies the trajectory-level relation
\begin{equation}
    \ln{\frac{P_{F}[\omega_{t}]}{P_{B}[\Theta\omega_{t}]}} = \frac{\Sigma_{t}[\omega_{t}]}{k_{B}}.
\end{equation}
After averaging over the forward paths, we have
\begin{equation}
    D(P_{F}\Vert P_{B}) = \int\mathcal{D}\omega_{t}P_{F}[\omega_{t}]\ln{\frac{P_{F}[\omega_{t}]}{P_{B}[\Theta\omega_{t}]}} = \frac{\langle\Sigma_t\rangle}{k_{B}}.
\end{equation}
Combining this identity with Eq.~\eqref{eq: task.DPI}, we obtain the fixed-time task entropy bound
\begin{equation}\label{eq: task.entropy.bound}
    \frac{\langle\Sigma_t\rangle}{k_B} \ge D_{\rm bin}\!\left(q_F(t)\middle\|q_B(t)\right),
\end{equation}
This inequality does not require the reverse task to be rare. The construction of $q_F(t)$ and $q_B(t)$ is given from the joint dynamics of the internal state, the work count, and the running maximum.

The bound in Eq.~(\ref{eq: task.entropy.bound}) also has a particularly transparent form when the reverse task is rare. If the forward task is completed with probability close to one, $q_F(t)\simeq 1$, while the reverse-task probability is small,
$q_B(t)\ll 1$, then
\begin{equation}
    D_{\rm bin}\!\left(q_F(t)\middle\|q_B(t)\right)
    \simeq \ln\frac{1}{q_B(t)}.
\end{equation}
Consequently, Eq.~(\ref{eq: task.entropy.bound}) implies
\begin{equation}
    \frac{\dot S_{\rm tot}t}{k_B}
    \gtrsim \ln\frac{1}{q_B(t)}, \quad \langle\Sigma_{t}\rangle = \dot{S}_{\mathrm{tot}}t,
\end{equation}
where $\dot S_{\rm tot}$ is the steady-state entropy-production rate. Thus, making the reverse task exponentially unlikely requires a correspondingly large total entropy production over the observation interval.

\textit{Inequality for heat engines.---}
We use the convention that $Q_{H}$ and $Q_{C}$ denote heat transferred into the system, while $W_{\mathrm{out}}>0$ denotes work extracted from the system. The first law is therefore
\begin{equation}
    W_{\mathrm{out}}= Q_{H}+Q_{C}-\Delta E.
\end{equation}
Together with $\Sigma_{t} = \Delta s_{\mathrm{sys}} -\frac{Q_{H}}{T_{H}}-\frac{Q_{C}}{T_{C}}$, this gives
\begin{equation}
    W_{\mathrm{out}} = \eta_{C} Q_{H} -T_{C}\Sigma_{t} +T_{C}\Delta s_{\mathrm{sys}} -\Delta E,
\end{equation}
where $\eta_{C}=1-T_{C}/T_{H}$. Taking the forward average and applying the task entropy bound yields
\begin{equation}\label{eq:engine_boundary}
    \begin{split}
        \langle W_{\mathrm{out}}\rangle & \leq \eta_{C}\langle Q_{H}\rangle - k_{B}T_{C}D_{\mathrm{bin}}(q_{F}(t)\Vert q_{B}(t)) \\ 
        & \qquad + T_{C}\langle\Delta s_{\mathrm{sys}}\rangle - \langle\Delta E\rangle
    \end{split}
\end{equation}
For stationary operation the average boundary terms vanish, $\langle\Delta s_{\mathrm{sys}}\rangle=\langle\Delta E\rangle=0$. Defining $\eta(t)=\langle W_{\mathrm{out}}\rangle/\langle Q_{H}\rangle$ for $\langle Q_{H}\rangle>0$, we obtain
\begin{equation}\label{eq:engine_efficiency}
    \eta(t) \leq \eta_{\mathrm{task}} = \eta_{C} - \frac{k_{B}T_{C}}{\langle Q_{H}\rangle}D_{\mathrm{bin}}\big(q_{F}^{\mathrm{eng}}(t)\Vert q_{B}^{\mathrm{eng}}(t)\big).
\end{equation}

\textit{Inequality for refrigerators.---}
For refrigerator operation, the first law and the entropy balance are $W_{\mathrm{in}} + Q_{C} + Q_{H} = \Delta E$ and $\Sigma_{t} = \Delta s_{\mathrm{sys}} - \frac{Q_{H}}{T_{H}} - \frac{Q_{C}}{T_{C}}$. Eliminating $Q_{H}$ gives
\begin{equation}\label{eq:fridge_identity}
    W_{\mathrm{in}} = \frac{Q_{C}}{\mathrm{COP}_{C}} + T_{H}\Sigma_{t} - T_{H}\Delta s_{\mathrm{sys}} + \Delta E,
\end{equation}
where $\mathrm{COP}_{C} = \frac{T_{C}}{T_{H}-T_{C}}$. Averaging Eq.~\eqref{eq:fridge_identity} and applying the task
entropy bound for the refrigerator task gives
\begin{equation}\label{eq:fridge_boundary}
    \begin{split}
        \langle W_{\mathrm{in}}\rangle &\geq \frac{Q_{C}}{\mathrm{COP}_{C}} + k_{B}T_{H}D_{\mathrm{bin}}(q_{F}(t)\Vert q_{B}(t)) \\ 
        & \qquad - T_{H}\langle\Delta s_{\mathrm{sys}}\rangle + \langle\Delta E\rangle.
    \end{split}
\end{equation}
For stationary operation with vanishing average boundary terms, and defining $\mathrm{COP}(t) = \langle Q_{C}\rangle/\langle W_{\mathrm{in}}\rangle$, we obtain
\begin{equation}\label{eq:fridge_COP}
    \frac{1}{\mathrm{COP}(t)} \geq \frac{1}{\mathrm{COP}_{\mathrm{task}}} = \frac{1}{\mathrm{COP}_{C}} + \frac{k_{B}T_{H}}{Q_{C}}D_{\mathrm{bin}}\big(q_{F}^{\mathrm{ref}}(t) \Vert q_{B}^{\mathrm{ref}}(t)\big).
\end{equation}

\begin{figure}
    \centering
    \includegraphics[width=1\linewidth]{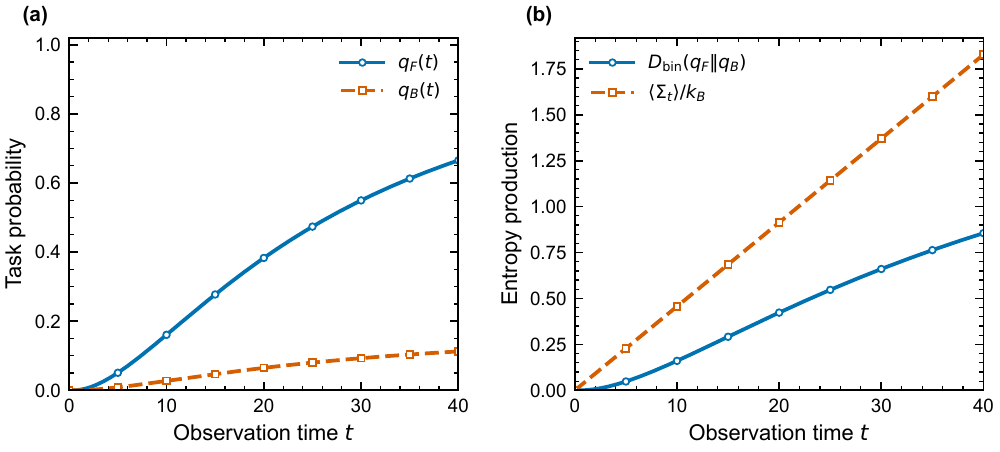}
    \caption{Task probabilities and coarse-grained entropy production for the forward and strictly time-reversed tasks at threshold $N=2$ in the heat engine regime. (a) Forward-task probability $q_F(t)$ and reverse-task probability $q_B(t)$. (b) Binary relative entropy $D_{\mathrm{bin}}(q_{F}\Vert q_{B})$ and the mean total entropy production $\langle \Sigma_{t} \rangle/k_{B}$. The curves satisfy the data-processing bound $D_{\mathrm{bin}}(q_{F}\Vert q_{B}) \leq \langle \Sigma_{t} \rangle/k_{B}$. }
    \label{fig: task.entropy}
\end{figure}

\textit{Results.---}
We first consider the probabilities of a forward task and its strictly time-reversed counterpart with the total entropy production. The task event $A_{N}(t)$ occurs when the running maximum of the net work count reaches the threshold $N$, and the corresponding reverse event is $\Theta A_{N}(t)$. Fig.~\ref{fig: task.entropy}~(a) shows the forward-task probability $q_{F}(t)=P_{F}[A_{N}(t)]$ and reverse-task probability $q_{B}(t) = P_{B}[\Theta A_{N}(t)]$ for $N=2$. This figure presents the results for heat engine operation. Both probabilities start at zero because the initial count and running maximum are below the threshold. As the observation time increases, for the forward task in this engine regime, completing the forward work-count task is more probable than realizing its strictly time-reversed event, i.e., $q_{F}(t)\geq q_{B}(t)$. Fig.~\ref{fig: task.entropy}~(b) compares the resulting binary relative entropy with the mean total entropy production. For the stationary three-state model, the mean total entropy production is 
\begin{equation}
    \frac{\langle\Sigma_{t}\rangle}{k_{B}} = J_{\mathrm{cyc}}\mathcal{A}t,
\end{equation}
where $J_{\mathrm{cyc}}$ is the steady-state cycle current \cite{Schnakenberg1976rmp, Seifert2012rpp}, measured in completed forward cycles per unit time, and $\mathcal{A}$ is the dimensionless cycle affinity. For the symmetric work channel used here, the affinity is 
\begin{equation}
    \mathcal{A} = \frac{\epsilon_{C}}{k_{B}T_{C}} - \frac{\epsilon_{H}}{k_{B}T_{H}}.
\end{equation}
The internal state is initialized in its stationary distribution. Unless otherwise stated, the parameters are $T_{H}=2$, $T_{C}=1$, $\epsilon_{H}=4$, $W_{\mathrm{cycle}}=1$, $\epsilon_{C}=\epsilon_{H}-W_{\mathrm{cycle}}=3$, $k_{B}=1$, and $\gamma_{H}=\gamma_{C}=\gamma_{W}=1$. The thermal transition rates follow the symmetric-exponential prescription $k_{\uparrow,\downarrow}^{(\alpha)} = \gamma\exp[\mp\epsilon/(2k_BT_{\alpha})],\quad \alpha\in\{H, C\}$. We use a work-count cutoff $K=80$ and evolve over $0\leq t\leq 40$.

\begin{figure}
    \centering
    \includegraphics[width=1\linewidth]{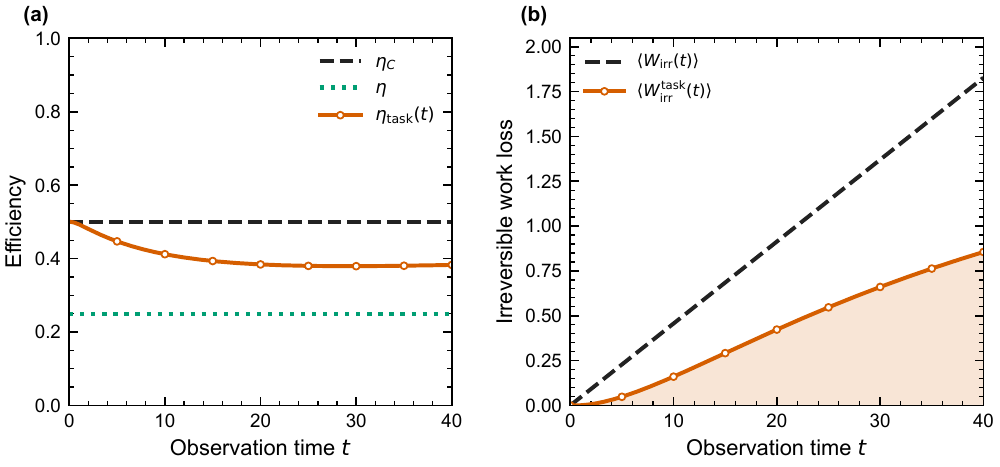}
    \caption{Task-dependent efficiency bound and irreversible work loss in the heat engine regime. (a) Carnot efficiency $\eta_{C}$, actual efficiency $\eta$, and task-dependent efficient bound $\eta_{\mathrm{task}}(t)$. The bound satisfies $\eta \leq \eta_{\mathrm{task}} \leq \eta_{C}$. (b) Actual irreversible loss $\langle W_{\mathrm{irr}}(t)\rangle  = (\eta_{C}-\eta)\langle Q_{H}(t)\rangle$ and the task-resolved lower bound $\langle W_{\mathrm{irr}}^{\mathrm{task}}(t)\rangle = k_{B}T_{C}D_{\mathrm{bin}}(q_{F}\Vert q_{B})$. }
    \label{fig: engine.loss}
\end{figure}

We next examine how much of the engine's irreversible work loss is resolved by the binary task description. Fig.~\ref{fig: engine.loss}~(a) compares this bound with the actual efficiency and the Carnot efficient. For the stationary model consider here, the actual efficient is time independent, 
\begin{equation}
    \eta = \frac{\langle W_{\mathrm{out}}(t)\rangle}{\langle Q_{H}(t)\rangle} = \frac{W_{\mathrm{cycle}}J_{\mathrm{cyc}}t}{\epsilon_{H}J_{\mathbf{cyc}}t} = \frac{W_{\mathrm{cycle}}}{\epsilon_{H}}.
\end{equation}
In contrast, the task-dependent bound $\eta_{\mathrm{task}}$ varies with the observation time through the binary relative entropy and the mean absorbed heat. At very short times, task completion is unlikely for both the forward and reverse processes. Consequently, the binary task contains little distinguishing information and the task-dependent bound lies close to the Carnot value. As time increases, the efficiency bound is tightened due to the forward and reverse task probabilities being more distinguishable. Fig.~\ref{fig: engine.loss}~(b) presents the same results in energetic form. The total irreversible work loss is compared with the task-resolved lower bound obtained from the binary relative entropy. The shaded region represents the portion of the total loss that is not resolved by the binary task. The central numerical check for Eq.~(\ref{eq:engine_efficiency}) is satisfied.

\begin{figure}
    \centering
    \includegraphics[width=1\linewidth]{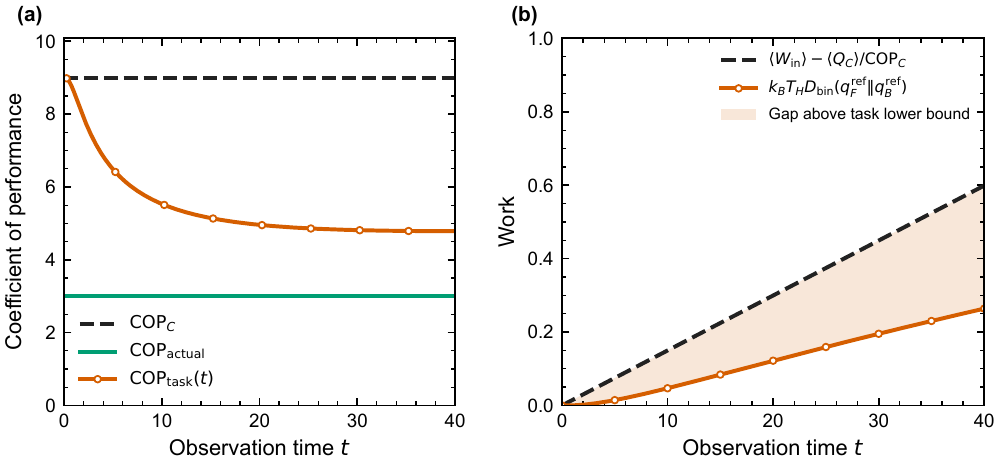}
    \caption{Task-dependent lower bound on the additional input work of a refrigerator. The parameters are chosen in the refrigerator regime, with $T_{H}>T_{C}$ and negative cycle current under the forward-cycle convention. (a) Actual coefficient of performance $\mathrm{COP}_{\mathrm{actual}} = \frac{\langle Q_{C}\rangle}{\langle W_{\mathrm{in}}\rangle}$, task-bound $\mathrm{COP}_{\mathrm{task}}(t)$ and Carnot coefficient of performance $\mathrm{COP}_{C}$. (b) Excess input work above the reversible Carnot coefficient, $\langle W_{\mathrm{in}}\rangle - \frac{Q_{C}}{\mathrm{COP}_{C}}$, compared with the task-dependent lower bound $k_{B}T_{H}D_{\mathrm{bin}}(q_{F}\Vert q_B)$. }
    \label{fig: refrigerator.lower}
\end{figure}

\begin{figure}
    \centering
    \includegraphics[width=1\linewidth]{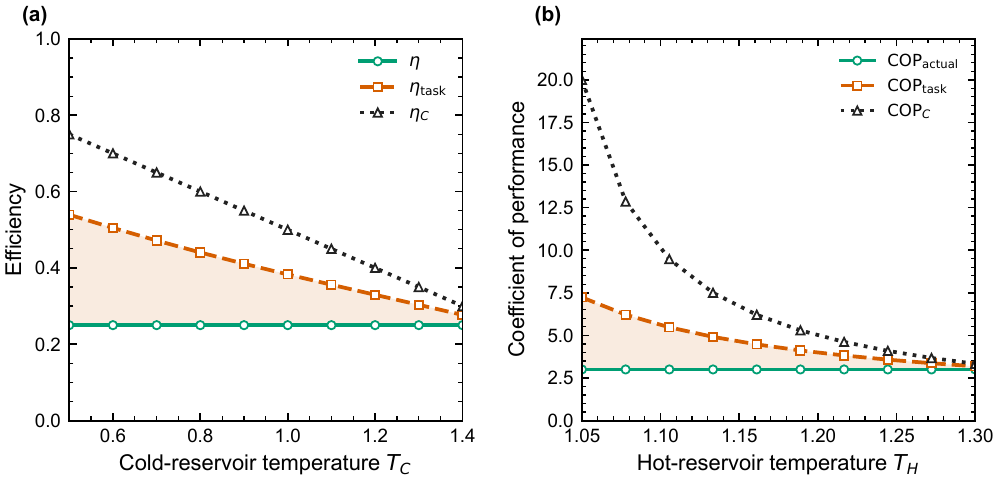}
    \caption{Temperature dependence of the actual and task-dependent performance bounds at the fixed observation time. The task probabilities are calculated using threshold $N=2$. (a) Heat engine regime: actual efficiency $\eta$, task-dependent efficiency bound $\eta_{\mathrm{task}}$, and Carnot efficiency $\eta_{C}$ versus cold-reservoir temperature $T_{C}$. The shaded region masks the gap between $\eta$ and $\eta_{\mathrm{task}}$. (b) Refrigerator regime: actual coefficient of performance $\mathrm{COP}_{\mathrm{actual}}$, task-dependent upper bound $\mathrm{COP}_{\mathrm{task}}$, and Carnot coefficient of performance $\mathrm{COP}_{C}$ versus heat-reservoir temperature $T_{H}$. }
    \label{fig: temperature.dependent}
\end{figure}

Similarly, the results of the refrigeration regime is considered in Fig.~\ref{fig: refrigerator.lower}. Fig.~\ref{fig: refrigerator.lower}~(a) compares the actual $\mathrm{COP}_{\mathrm{actual}}$, Carnot $\mathrm{COP}_{C}$ and task-bound $\mathrm{COP}_{\mathrm{task}}$ reversible coefficients of performance. As expected for an irreversible device, the actual coefficient of performance remains below the Carnot value. Fig.~\ref{fig: refrigerator.lower}~(b) presents the corresponding
energetic form of the task-dependent bound. The excess input work above the reversible Carnot requirement, $\langle W_{\mathrm{in}}\rangle-\langle Q_C\rangle/\mathrm{COP}_C$ is compared with the task-dependent lower bound $k_BT_H D_{\mathrm{bin}}(q_F\Vert q_B)$. The parameters are $T_H=2.0$ and $T_C=1.8$, with the
standard parameter set specified above. The numerical results satisfy Eq.~(\ref{eq:fridge_COP}) showing that the binary task provides a lower bound on the input work in excess of the reversible requirement.

Finally, we examine the temperature dependence of the task-dependent performance bounds at the fixed observation time $t=40$. In Fig.~\ref{fig: temperature.dependent}~(a), we fix $T_H=2$ and vary $T_C$ over the heat-engine temperature range used in the sweep. We plot the actual efficiency $\eta$, the task-dependent bound $\eta_{\mathrm{task}}$, and the Carnot efficiency $\eta_C$; the shaded region marks the gap between $\eta$ and $\eta_{\mathrm{task}}$. In Fig.~\ref{fig: temperature.dependent}~(b), we fix $T_C=1$ and vary $T_H$ over the refrigerator candidate range. The panel shows $\mathrm{COP}_{\mathrm{actual}}$, the task-dependent upper bound $\mathrm{COP}_{\mathrm{task}}$, and the Carnot value $\mathrm{COP}_C$.  The plotted
bounds are evaluated from $D_{\mathrm{bin}}(q_F\Vert q_B)$ at the final observation time. The results satisfy $\eta\leq\eta_{\mathrm{task}}\leq\eta_C$ in the engine regime and $\mathrm{COP}_{\mathrm{actual}}\leq\mathrm{COP}_{\mathrm{task}} \leq\mathrm{COP}_C$ in the refrigerator regime.

{\it Experimental outlook.---}
A superconducting circuit offers a route to testing these bounds, building on real-time monitoring of quantum jumps in a three-level artificial atom \cite{Garrahan2018pra, Minev2019nature}. A three-level maser design could couple two transitions to engineered hot and cold reservoirs and use the third as a monitored work channel with symmetric transition rates \cite{Vega2017rmp, Kalaee2021pre}. Repeated runs under the prescribed forward and reverse preparations would yield the work-count records needed to estimate the forward threshold-crossing and reverse terminal-drawdown probabilities. The resulting bounds could be tested against independently estimated mean heat and work exchanges.

{\it Conclusion.---}
We developed a framework linking finite-time task statistics to thermodynamic performance. The exact time reverse of reaching a work-count threshold is a terminal drawdown from the running maximum of the reversed count. By the data-processing inequality, the binary relative entropy of the forward and reverse success probabilities bounds mean total entropy production from below. When mean changes in system energy and entropy vanish, this bound sets minimum work losses for heat engines and minimum excess work inputs for refrigerators relative to reversible operation.

A three-level stochastic system illustrates the framework. For the parameters studied, the efficiency bound improves on the Carnot bound most at intermediate observation times. Once the two success probabilities are known, the entropy production bound requires no reconstruction of full trajectory distributions.

\begin{acknowledgments}
  K.C. was funded by the National Natural Science Foundation (Grants No.~12605047), Basic Research Program of Jiangsu (Grant No.~BK20250886) and Basic Research Program of Yangzhou (Grant No.~YZ2025132).  J.W. was supported by the National Natural Science Foundation of China (Grant No. 11875047).
\end{acknowledgments}

\bibliography{reference}

\begin{widetext}
    
\section*{Supplemental Material}

\begin{center}
\textbf{Time-Reversed Task Events and Thermodynamic Bounds for Stochastic Thermal Machines}
\end{center}

This Supplemental Material provides details of the three-state model, the exact time-reversed task event, the thermodynamic derivations of the engine and refrigerator bounds, and the numerical method used to evaluate the task probabilities.

\section{Three-state model and thermodynamic conventions}

We consider a three-state system with energies
\begin{equation}
E_1=0,\qquad E_2=\epsilon_C,\qquad E_3=\epsilon_H,
\label{eq:energies}
\end{equation}
where $0<\epsilon_C<\epsilon_H$. The transition $1\leftrightarrow3$ is coupled to the hot reservoir at temperature $T_H$, and $1\leftrightarrow2$ is coupled to the cold reservoir at temperature $T_C$. The transition $2\leftrightarrow3$ is coupled to an ideal work channel. We define
\begin{equation}
W_{\mathrm{cycle}}=\epsilon_H-\epsilon_C.
\label{eq:W_cycle}
\end{equation}

Heat is positive when it enters the system from a thermal reservoir. Work output is positive when extracted from the system; work input is positive when supplied to it. Thus, for engine operation,
\begin{equation}
\Delta E=Q_H+Q_C-W_{\mathrm{out}},
\label{eq:first_law_engine}
\end{equation}
whereas for refrigerator operation,
\begin{equation}
\Delta E=Q_H+Q_C+W_{\mathrm{in}}.
\label{eq:first_law_fridge}
\end{equation}

For $\alpha\in\{H,C\}$, the thermal rates obey local detailed balance,
\begin{equation}
\frac{k^{(\alpha)}_{i\to j}}{k^{(\alpha)}_{j\to i}}
=
\exp[-\beta_\alpha(E_j-E_i)],
\qquad
\beta_\alpha=(k_BT_\alpha)^{-1}.
\label{eq:ldb}
\end{equation}
For the numerical examples, we use the symmetric-exponential parametrization
\begin{equation}
k^{(\alpha)}_{\mathrm{up}}
=
\gamma_\alpha
\exp\!\left[-\frac{\epsilon_\alpha}{2k_BT_\alpha}\right],
\qquad
k^{(\alpha)}_{\mathrm{down}}
=
\gamma_\alpha
\exp\!\left[+\frac{\epsilon_\alpha}{2k_BT_\alpha}\right],
\label{eq:sym_rates}
\end{equation}
where $\epsilon_H=E_3-E_1$ and $\epsilon_C=E_2-E_1$. The work-channel rates are denoted by $k^{(W)}_{2\to3}$ and $k^{(W)}_{3\to2}$. For the symmetric work channel considered here,
\begin{equation}
k^{(W)}_{2\to3}=k^{(W)}_{3\to2}.
\label{eq:sym_work}
\end{equation}

With the column-vector convention, the population vector $\boldsymbol p=(p_1,p_2,p_3)^{\mathsf T}$ satisfies
\begin{equation}
\dot{\boldsymbol p}=L\boldsymbol p,
\label{eq:master_eq}
\end{equation}
with
\begin{equation}
L=
\begin{pmatrix}
-(k^{(C)}_{1\to2}+k^{(H)}_{1\to3})
& k^{(C)}_{2\to1}
& k^{(H)}_{3\to1}\\
k^{(C)}_{1\to2}
&-(k^{(C)}_{2\to1}+k^{(W)}_{2\to3})
&k^{(W)}_{3\to2}\\
k^{(H)}_{1\to3}
&k^{(W)}_{2\to3}
&-(k^{(H)}_{3\to1}+k^{(W)}_{3\to2})
\end{pmatrix}.
\label{eq:L_matrix}
\end{equation}
The stationary state satisfies
\begin{equation}
L\boldsymbol p_{\mathrm{ss}}=0,
\qquad
\sum_i p_i^{\mathrm{ss}}=1.
\label{eq:steady_state}
\end{equation}

We assign the work-count increments
\begin{equation}
g(3\to2)=+1,\qquad g(2\to3)=-1,
\label{eq:work_increments}
\end{equation}
and zero increment to thermal transitions. Hence
\begin{equation}
n(t)=\sum_{\ell:t_\ell\leq t} g(x_{\ell^-}\to x_{\ell^+}),
\qquad
W_{\mathrm{out}}(t)=W_{\mathrm{cycle}}n(t).
\label{eq:n_t_def}
\end{equation}

\section{Stationary current and operating regimes}

We orient the cycle as
\begin{equation}
1\longrightarrow3\longrightarrow2\longrightarrow1.
\label{eq:cycle_orientation}
\end{equation}
The stationary cycle current is
\begin{equation}
J_{\mathrm{cyc}}
=
p_1^{\mathrm{ss}}k^{(H)}_{1\to3}
-
p_3^{\mathrm{ss}}k^{(H)}_{3\to1}.
\label{eq:J_cyc}
\end{equation}
The same net current is obtained on each edge when the current is defined consistently with the chosen cycle orientation.

For the symmetric work channel, the cycle affinity is
\begin{equation}
\mathcal A
=
\ln\frac{
k^{(H)}_{1\to3}
k^{(W)}_{3\to2}
k^{(C)}_{2\to1}
}{
k^{(H)}_{3\to1}
k^{(W)}_{2\to3}
k^{(C)}_{1\to2}
}
=
\frac{\epsilon_C}{k_BT_C}
-
\frac{\epsilon_H}{k_BT_H}.
\label{eq:affinity}
\end{equation}
The stationary entropy-production rate is
\begin{equation}
\dot S_{\mathrm{tot}}=k_BJ_{\mathrm{cyc}}\mathcal A.
\label{eq:entropy_prod_rate}
\end{equation}
The engine and refrigerator regimes are identified by the sign of $J_{\mathrm{cyc}}$ under this cycle convention. The stall point has $J_{\mathrm{cyc}}=0$.

\section{Exact time reversal of the threshold task}

For a trajectory observed over $[0,t]$, define the net work count $n(s)$ and its running maximum
\begin{equation}
m(s)=\max_{0\le u\le s}n(u).
\label{eq:running_max}
\end{equation}
The forward task event is
\begin{equation}
A_N(t)=\{m(t)\geq N\}.
\label{eq:forward_task}
\end{equation}

Under exact trajectory reversal, the reversed count, measured from zero at the beginning of the reversed trajectory, is
\begin{equation}
\bar n(r)=n(t-r)-n(t),
\qquad 0\le r\le t.
\label{eq:reversed_count}
\end{equation}
Consequently,
\begin{equation}
\max_{0\le r\le t}\bar n(r)
=
\max_{0\le s\le t}n(s)-n(t)
=
m(t)-n(t).
\label{eq:max_reversed}
\end{equation}
The exact time-reversed image of the forward event is therefore
\begin{equation}
\Theta A_N(t)
=
\left\{\max_{0\le r\le t}\bar n(r)\geq N\right\}
=
\{m(t)-n(t)\geq N\}.
\label{eq:reversed_task}
\end{equation}
This is a terminal drawdown event. It is generally different from the fixed-boundary first-passage event $\{\min_{0\le r\le t}\bar n(r)\leq-N\}$.

For the autonomous model initialized in its stationary distribution, the forward and reverse experiments have the same transition rates and initial state distribution. Their task events are nevertheless different: the reverse task is the time-reversed image $\Theta A_N(t)$, not a second application of the forward running-maximum criterion.

\section{Binary coarse-graining and entropy production}

Let $P_F[\omega_t]$ denote the forward path measure and let $P_B[\Theta\omega_t]$ denote the probability of the time-reversed path under the reverse experiment. The stochastic total entropy production is
\begin{equation}
\Sigma_t
=
\Delta s_{\mathrm{sys}}+\Delta s_{\mathrm{med}},
\label{eq:total_entropy}
\end{equation}
where
\begin{equation}
\Delta s_{\mathrm{sys}}
=
-k_B\ln p_t(x_t)+k_B\ln p_0(x_0)
\label{eq:entropy_sys}
\end{equation}
and
\begin{equation}
\Delta s_{\mathrm{med}}
=
-\frac{Q_H}{T_H}-\frac{Q_C}{T_C}.
\label{eq:entropy_med}
\end{equation}
Local detailed balance gives
\begin{equation}
\ln\frac{P_F[\omega_t]}
{P_B[\Theta\omega_t]}
=
\frac{\Sigma_t[\omega_t]}{k_B}.
\label{eq:path_ratio}
\end{equation}
Averaging over forward paths yields
\begin{equation}
D(P_F\Vert P_B\circ\Theta)
=
\frac{\langle\Sigma_t\rangle_F}{k_B}.
\label{eq:KL_divergence}
\end{equation}

Coarse-grain a trajectory to the binary indicator of the forward event:
\begin{equation}
Y_t(\omega_t)=
\begin{cases}
1,&\omega_t\in A_N(t),\\
0,&\omega_t\notin A_N(t).
\end{cases}
\label{eq:binary_indicator}
\end{equation}
The probabilities of outcome $1$ under the two path measures are
\begin{equation}
q_F(t)=P_F[A_N(t)],
\qquad
q_B(t)=P_B[\Theta A_N(t)].
\label{eq:probabilities}
\end{equation}
The data-processing inequality gives
\begin{equation}
\frac{\langle\Sigma_t\rangle_F}{k_B}
\geq
D_{\mathrm{bin}}\!\left(q_F(t)\Vert q_B(t)\right),
\label{eq:DP_inequality}
\end{equation}
where
\begin{equation}
D_{\mathrm{bin}}(p\Vert q)
=
p\ln\frac pq
+(1-p)\ln\frac{1-p}{1-q}.
\label{eq:bin_entropy}
\end{equation}
We use the continuous endpoint convention $0\ln 0=0$; if $q_F>0$ and $q_B=0$, the binary relative entropy diverges.

\subsection{Heat engine}

With $W_{\mathrm{out}}>0$ denoting work extracted from the system, the first law is
\begin{equation}
W_{\mathrm{out}}=Q_H+Q_C-\Delta E.
\label{eq:first_law_engine_2}
\end{equation}
Combining this with
\begin{equation}
\Sigma_t=\Delta s_{\mathrm{sys}}
-\frac{Q_H}{T_H}-\frac{Q_C}{T_C}
\label{eq:sigma_combined}
\end{equation}
gives the trajectory identity
\begin{equation}
W_{\mathrm{out}}
=
\eta_C Q_H
-T_C\Sigma_t
+T_C\Delta s_{\mathrm{sys}}
-\Delta E,
\qquad
\eta_C=1-\frac{T_C}{T_H}.
\label{eq:W_out_identity}
\end{equation}
Taking the forward average and applying the binary entropy bound,
\begin{equation}
\begin{aligned}
\langle W_{\mathrm{out}}\rangle
\leq{}&
\eta_C\langle Q_H\rangle
-k_BT_C D_{\mathrm{bin}}(q_F\Vert q_B)\\
&+T_C\langle\Delta s_{\mathrm{sys}}\rangle
-\langle\Delta E\rangle.
\end{aligned}
\label{eq:W_out_bound}
\end{equation}
For stationary initialization, $\langle\Delta s_{\mathrm{sys}}\rangle=0$. If the mean internal energy is unchanged, $\langle\Delta E\rangle=0$, and $\langle Q_H\rangle>0$, then
\begin{equation}
\eta
=
\frac{\langle W_{\mathrm{out}}\rangle}{\langle Q_H\rangle}
\leq
\eta_{\mathrm{task}}
=
\eta_C
-\frac{k_BT_C}{\langle Q_H\rangle}
D_{\mathrm{bin}}(q_F\Vert q_B).
\label{eq:eta_bound}
\end{equation}

\subsection{Refrigerator}

For refrigerator operation, $Q_C>0$ is the heat absorbed from the cold reservoir and $W_{\mathrm{in}}>0$ is the work supplied to the system. The first law is
\begin{equation}
\Delta E=Q_H+Q_C+W_{\mathrm{in}},
\label{eq:first_law_fridge_2}
\end{equation}
and the total entropy production is
\begin{equation}
\Sigma_t
=
\Delta s_{\mathrm{sys}}
-\frac{Q_H}{T_H}
-\frac{Q_C}{T_C}.
\label{eq:sigma_fridge}
\end{equation}
Eliminating $Q_H$ gives
\begin{equation}
W_{\mathrm{in}}
=
\frac{Q_C}{\mathrm{COP}_C}
+T_H\Sigma_t
-T_H\Delta s_{\mathrm{sys}}
+\Delta E,
\qquad
\mathrm{COP}_C=\frac{T_C}{T_H-T_C}.
\label{eq:W_in_identity}
\end{equation}
Averaging and applying the task entropy bound yields
\begin{equation}
\begin{aligned}
\langle W_{\mathrm{in}}\rangle
\geq{} \frac{\langle Q_C\rangle}{\mathrm{COP}_C}
+k_BT_H D_{\mathrm{bin}}(q_F\Vert q_B) -T_H\langle\Delta s_{\mathrm{sys}}\rangle
+\langle\Delta E\rangle.
\end{aligned}
\label{eq:W_in_bound}
\end{equation}
For stationary initialization with vanishing mean energy change,
\begin{equation}
\langle\Delta s_{\mathrm{sys}}\rangle=0,
\qquad
\langle\Delta E\rangle=0,
\label{eq:stationary_assumptions}
\end{equation}
and $\langle Q_C\rangle>0$, this becomes
\begin{equation}
\langle W_{\mathrm{in}}\rangle
-\frac{\langle Q_C\rangle}{\mathrm{COP}_C}
\geq
k_BT_H D_{\mathrm{bin}}(q_F\Vert q_B).
\label{eq:COP_bound_1}
\end{equation}

Defining the task-dependent COP upper bound by
\begin{equation}
\frac{1}{\mathrm{COP}_{\mathrm{task}}}
=
\frac{1}{\mathrm{COP}_C}
+
\frac{k_BT_H}{\langle Q_C\rangle}
D_{\mathrm{bin}}(q_F\Vert q_B),
\label{eq:COP_task_def}
\end{equation}
we obtain
\begin{equation}
\mathrm{COP}_{\mathrm{actual}}
\leq
\mathrm{COP}_{\mathrm{task}}
\leq
\mathrm{COP}_C.
\label{eq:COP_hierarchy}
\end{equation}

\section{Extended-state-space numerical method}

The task probabilities depend on the trajectory history through the running maximum. We therefore propagate the joint probability $p_{i,n,m}(t)$ of the internal state $i$, the current count $n$, and the running maximum
\begin{equation}
m=\max_{0\leq s\leq t}n(s).
\label{eq:running_max_num}
\end{equation}
The initial condition is
\begin{equation}
p_{i,n,m}(0)
=
p_i^{\mathrm{ss}}\,
\delta_{n,0}\delta_{m,0}.
\label{eq:initial_condition}
\end{equation}

Thermal jumps leave $(n,m)$ unchanged. A work jump $3\to2$ updates
\begin{equation}
(n,m)\longrightarrow(n+1,\max\{m,n+1\}),
\label{eq:jump_32}
\end{equation}
whereas a jump $2\to3$ updates
\begin{equation}
(n,m)\longrightarrow(n-1,m).
\label{eq:jump_23}
\end{equation}
Thus, the extended process is a continuous-time Markov process on $(i,n,m)$.

At each observation time, the forward and reverse task probabilities are obtained from the same joint distribution:
\begin{equation}
q_F(t)=\sum_{i,n,m:\,m\geq N}p_{i,n,m}(t),
\label{eq:q_F_num}
\end{equation}
\begin{equation}
q_B(t)=\sum_{i,n,m:\,m-n\geq N}p_{i,n,m}(t).
\label{eq:q_B_num}
\end{equation}
The second expression evaluates the terminal drawdown event and is not a fixed negative-threshold first-passage probability.

\begin{figure}
    \centering
    \includegraphics[width=1\linewidth]{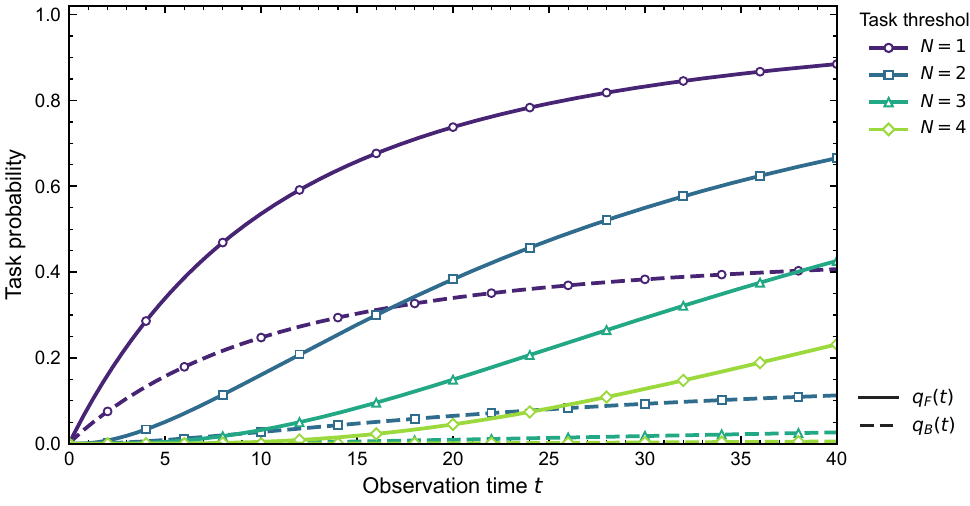}
    \caption{ Forward and reverse task probabilities, $q_{F}(t)$ and $q_{B}(t)$, as functions of the observation time $t$. The task is defined by a threshold $N$ in the cycle count.}
    \label{fig: task.probability}
\end{figure}

Fig.~\ref{fig: task.probability} compares the probabilities of the forward task and its time-reversed event. Unless stated otherwise, all results are obtained from a single propagation with count cutoff $K = 80$. The stationary distribution is first obtained from $Lp^{\mathrm{ss}} = 0$, and the resulting state is used as the initial condition. The dynamics is then lifted to the extended state space $(i, n, m)$, where $i \in \{1, 2, 3\}$ is the internal state, $n = n(t)$ is the current net work count, and $m = \max_{0 \leq s \leq t}n(s)$ is the running maximum. The retained states satisfy
\begin{equation}
    n\in[-K, K], \quad m\in[0, K], \quad m\geq\max{(n,0)}.
\end{equation}

\begin{figure}
    \centering
    \includegraphics[width=1\linewidth]{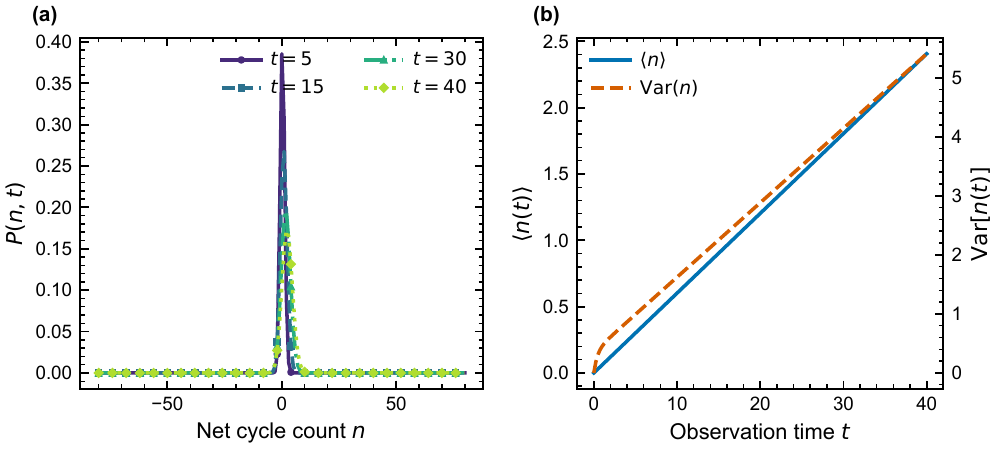}
    \caption{Net cycle-count statistics as a function of observation time. (a) The marginal distribution $P(n,t)$, obtained by summing    $p_{i,n,m}(t)$ over the internal state $i$ and running maximum $m$, broadens and shifts toward positive count values as time increases. (b) The mean count $\langle n(t)\rangle$ and its steady-state prediction $\mathcal{J}t$, together with the count variance $\mathrm{Var}[n(t)]$. The mean follows the linear prediction set by the net cycle current $\mathcal{J}$, while the variance grows approximately linearly at long times.}
    \label{fig: cycle.count}
\end{figure}

Fig.~\ref{fig: cycle.count} describes the statistics of the net work count underlying the task probabilities in Fig.~\ref{fig: task.probability}. The marginal count distribution
\begin{equation}
    P(n,t)=\sum_{i,m}p_{i,n,m}(t)
\end{equation}
shows how likely each net count is at time $t$. As the observation time increases, $P(n,t)$ shifts toward positive values and broadens [see Fig.~\ref{fig: cycle.count}~(a)]. The shift reflects the positive stationary cycle current, whereas the broadening results from stochastic fluctuations in the number of forward and backward work-channel transitions.

The first two moments make these two effects explicit:
\begin{equation}
    \langle n(t)\rangle=\sum_n nP(n,t),
\end{equation}
and 
\begin{equation}
    \mathrm{Var}[n(t)]=\sum_n n^2P(n,t)-\langle n(t)\rangle^2.
\end{equation}
The mean count grows approximately linearly with time and agrees with the stationary-current prediction
\begin{equation}
    \langle n(t) \rangle \simeq J_{\mathrm{cyc}}t
\end{equation}
with $J_{\rm cyc} = p^{\rm ss}_1 k_{1\to3} - p^{\rm ss}_3 k_{3\to1}$. This agreement verifies that the extended-state propagation reproduces the stationary current of the underlying three-state process. The variance also grows approximately linearly over the displayed time interval, consistent with diffusive growth of current fluctuations.

The positive drift favors forward threshold crossings, while the diffusive broadening allows trajectories to move back from previously reached positive counts. Thus, Fig.~\ref{fig: cycle.count} explains why the forward task probability tends to exceed the reverse probability in Fig.~\ref{fig: task.probability}, but the marginal distribution $P(n,t)$ alone is insufficient to determine the reverse task. The latter depends on the largest count reached during the entire trajectory, not only on the final value of $n$. This history dependence is resolved in Fig.~\ref{fig: joint.retreat}.

\begin{figure}
    \centering
    \includegraphics[width=1\linewidth]{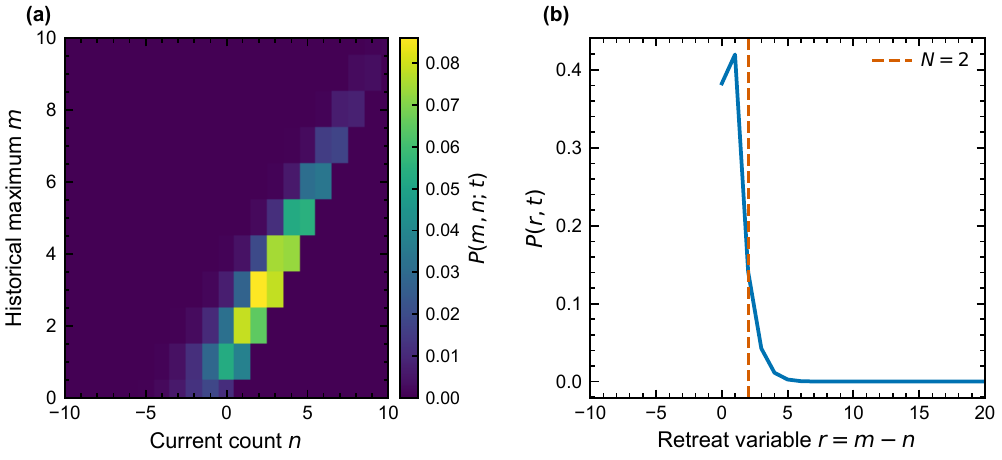}
    \caption{Joint and drawdown statistics of the net cycle count at $t=40$. (a) The joint probability $P(m,n;t)$ of the current net cycle count $n$ and its running maximum $m=\max_{0\le s\le t}n(s)$, obtained by summing $p_{i,n,m}(t)$ over the internal state $i$. (b) The corresponding drawdown distribution $P(r,t)$ for $r=m-n$, obtained by summing $P(m,n;t)$ along the diagonals $m-n=r$. The dashed line marks the task threshold $r=N=2$. The reverse-task probability is the integrated tail $q_B(t)=\sum_{r\ge N}P(r,t)$, whereas the forward-task probability is obtained from the running-maximum tail $q_F(t)=\sum_{m\ge N,n}P(m,n;t)$.}
    \label{fig: joint.retreat}
\end{figure}

Fig.~\ref{fig: joint.retreat} retains the path-history information by considering the joint distribution of the final count and the running maximum at $t=40$,
\begin{equation}
    P(m,n;t)=\sum_i p_{i,n,m}(t).
\end{equation}
Here, the variable $n$ denotes the net cycle count at the observation time, whereas $m=\max_{0\le s\le t}n(s)$ records the largest count reached at any earlier time, including the final time.
Because the process starts from $n(0)=0$, the allowed support satisfies $m\geq \max\{n,0\}$. The condition $m\geq n$ follows directly from the definition of a running maximum, while $m\geq0$ follows from the initial count. The probability density is therefore confined to a one-sided region of the $(n,m)$ plane. The weight extending into $m>n$ represents trajectories that have retreated from a previously larger count. For example, a trajectory can reach a positive maximum and subsequently end at a smaller, or even negative, value of $n$. The concentration of the distribution in the displayed range indicates the counts and running maxima that are most probable at $t=40$; the location of its maximum should not be interpreted as a deterministic trajectory or as the mean value of either variable.

The left panel is directly related to the results shown previously. Marginalizing over the running maximum gives the count distribution $P(n,t)=\sum_m P(m,n;t)$, which is the quantity plotted in Fig.~\ref{fig: cycle.count}. Likewise, the forward task probability is the probability that the running maximum has reached the threshold:
\begin{equation}
    q_F(t)=\Pr[m\geq N] = \sum_{m\geq N}\sum_n P(m,n;t).    
\end{equation}
Consequently, the forward-task probability in Fig.~\ref{fig: task.probability} is obtained from the upper-$m$ tail of the joint distribution.

\begin{figure}
    \centering
    \includegraphics[width=\textwidth]{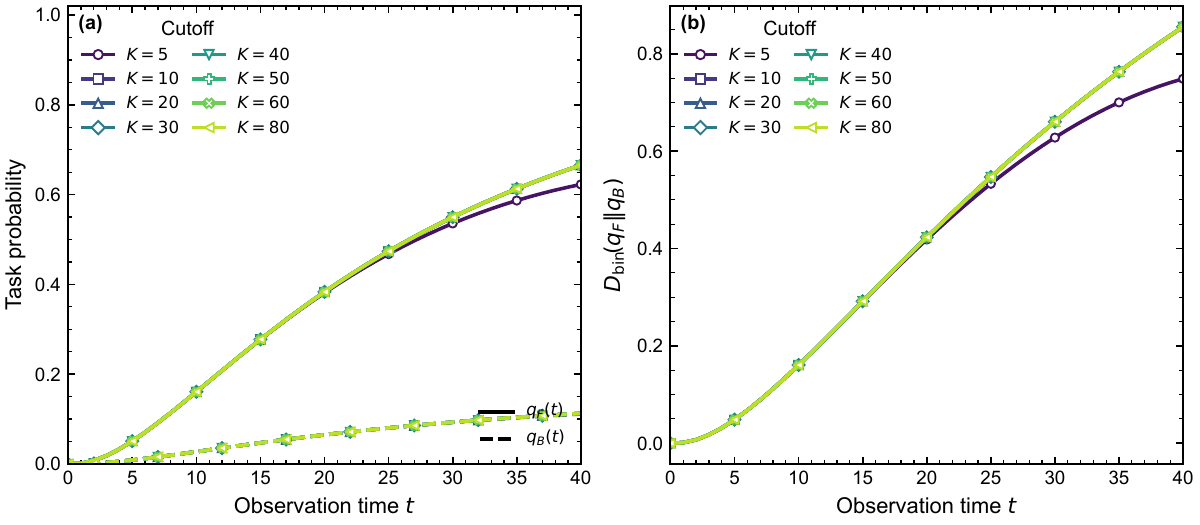}
    \caption{
    Convergence of task probabilities and binary relative entropy
    with respect to the work-count cutoff.
    (a) Forward task probability $q_F(t)$ (solid lines) and reverse
    task probability $q_B(t)$ (dashed lines) for cutoffs
    $K=5,10,20,30,40,50,60,$ and $80$.
    (b) Binary relative entropy
    $D_{\mathrm{bin}}(q_F\Vert q_B)$ for the same cutoffs.
    Colors and marker shapes distinguish $K$.
    All curves use task threshold $N=2$, the model parameters
    specified in the main text, and 41 equally spaced observation
    times over $0\leq t\leq 40$.
    The $K=80$ result is a numerical reference for the cutoff
    comparison, not an exact solution.
    }
    \label{fig:cutoff_convergence}
\end{figure}

The extended-state master equation is propagated on a finite domain
with work-count cutoff $K$. To assess the effect of this truncation,
we repeated the calculation for
$K=5,10,20,30,40,50,60,$ and $80$, keeping all physical parameters
and the task threshold $N=2$ fixed. For each cutoff, we computed the
forward and reverse task probabilities, $q_F(t)$ and $q_B(t)$, and
their binary relative entropy,
$D_{\mathrm{bin}}(q_F\Vert q_B)$, over $0\leq t\leq 40$ using
41 equally spaced observation times.

Fig.~\ref{fig:cutoff_convergence} shows these observables for all
cutoffs. The largest cutoff is used as a numerical reference, not as an exact
solution. The retained probability is not renormalized during
propagation; its deficit from unity is also monitored as a diagnostic
of probability loss at the finite-domain boundary. Since a small
normalization deficit alone does not establish convergence of a
particular observable, convergence is assessed directly from
$q_F$, $q_B$, and $D_{\mathrm{bin}}$.

\end{widetext}
\end{document}